\documentclass[letterpaper,preprintnumbers,prd,twocolumn,nofootinbib,nobibnotes,showpacs]{revtex4}
\usepackage{amsfonts}
\usepackage{mathrsfs}
\usepackage{epsfig}
\usepackage{graphicx}%
\usepackage{dcolumn}
\usepackage{amsmath}
\usepackage{color}
\usepackage{color}
\makeatletter
\def\btt#1{\texttt{\@backslashchar#1}}%
\DeclareRobustCommand\bblash{\btt{\@backslashchar}}%
\makeatother
\begin{document}

\title{Black hole and cosmos with multiple horizons and multiple singularities in vector-tensor theories}
\author{Changjun Gao}\email{gaocj@bao.ac.cn} \affiliation{ Key Laboratory of Computational Astrophysics, National Astronomical Observatories, Chinese
Academy of Sciences, Beijing 100012, China}
\affiliation{School of Astronomy and Space Sciences, University of Chinese Academy of Sciences,
No. 19A, Yuquan Road, Beijing 100049, China}\author{Youjun Lu}\email{luyj@nao.cas.cn}\affiliation{ Key Laboratory of Computational Astrophysics, National Astronomical Observatories, Chinese
Academy of Sciences, Beijing 100012, China}
\affiliation{School of Astronomy and Space Sciences, University of Chinese Academy of Sciences,
No. 19A, Yuquan Road, Beijing 100049, China}
\author{Shuang Yu}\email{yushuang@nao.cas.cn}\affiliation{ Key Laboratory of Computational Astrophysics, National Astronomical Observatories, Chinese
Academy of Sciences, Beijing 100012, China}
\affiliation{School of Astronomy and Space Sciences, University of Chinese Academy of Sciences,
No. 19A, Yuquan Road, Beijing 100049, China}
\author{You-Gen Shen}
\email{ygshen@center.shao.ac.cn} \affiliation{Shanghai
Astronomical Observatory, Chinese Academy of Sciences, Shanghai
200030, China}

\date{\today}

\begin{abstract}
 A stationary and spherically symmetric black hole (e.g., Reissner-Nordstr$\ddot{\textrm{o}}$m black hole or Kerr-Newman black hole) has at most one singularity and two horizons. One horizon is the outer event horizon and the other is the inner Cauchy horizon. Can we construct static and spherically symmetric black hole solutions with $N$ horizons and $M$ singularities?  De Sitter cosmos has only one apparent horizon. Can we construct cosmos solutions with $N$ horizons?  In this article, we present the static and spherically symmetric black hole and cosmos solutions with $N$ horizons and $M$ singularities in the vector-tensor theories. Following these motivations, we also construct the black hole solutions with a firewall. The deviation of these black hole solutions from the usual ones can be potentially tested by future measurements of gravitational waves or black hole continuum spectrum.
\end{abstract}



\maketitle

\section{Introduction}
Observations of gravitational waves by advanced LIGO \cite{abb:2016} open a new window for probing
black hole physics.  Any deviation of black holes in alternative theories of gravity from those in the Einstein's general relativity can be potentially tested by precise measurements of gravitational waves. In this new age of gravity studies, it is of great importance to
seek for new black hole solutions in alternative theories of gravity.

It is well known that the Schwarzschild black hole has one event horizon. The Reissner-Nordstr$\ddot{\textrm{o}}$m black hole (the Kerr black hole and the Kerr-Newman black hole) has two horizons. The Reissner-Nordstr$\ddot{\textrm{o}}$m-de Sitter (the Kerr-de Sitter and the Kerr-Newman-de Sitter black hole) black hole has three horizons. Note that all these black holes have a singularity in its center. Some interesting questions one would ask are as follows: 1) do black hole solutions with $N$ horizons exist? 2) can black holes with multiple singularities be constructed?
In fact, regular multi-horizon black holes in General Relativity, Lovelock theory and  modified gravity with non-linear
electrodynamics have been presented in Ref.~\cite{noj:2017}.
The spacetime with multiple singularities and multiple horizons is highly likely to be produced in the merger process of multiple black holes. But it is rather involved to obtain the analytic solutions describing this process. A few of the studies are as follows.

The Majumdar-Papapetrou solution describes multiple extreme Reissner-Nordstr$\ddot{\textrm{o}}$m black holes \cite{har:1972}. The Kastor-Traschen solution \cite{kas:1993,bri:1994} generalized the Majumdar-Papapetrou solution into the background of de Sitter universe. So it can describe the collisions of several black holes in asymptotically de Sitter space-time.  It is worth noting that the solution is the first exact solution that describes black holes collisions.  The dilaton version of the Kastor-Traschen solution is given in \cite{gao:2004} and the spinning version of the dilatonic Kastor-Traschen solution is given in \cite{shi:1999}. Within five-dimensional supergravity, Ref. \cite{beh:2003} provides examples of multi-centered charged
black holes in asymptotic de Sitter space. Using the compactification of intersecting
brane solution in higher dimensional unified
theory, a time-dependent cosmological black hole
system was found in \cite{mae:2009}. The global picture
of dynamical solution describes a multi-black
hole system in the expanding Universe filled by
``stiff matter'' is  clarified in \cite{mae:2010}. Finally,
the multiple black hole solution of general relativity
with a scalar field and two Maxwell-type U$(1)$ fields
is found  in \cite{gib:2010}.

In this paper, we present some toy models for the black holes spacetime with multiple horizons and multiple singularities.  As toy models, the solutions are all static and spherically symmetric. Therefore, the multiple horizons are concentric spheres and the singularities consists of one singular point and
multi-singular spheres. These spacetimes can be viewed as the generalization of Reissner-Nordstr$\ddot{\textrm{o}}$m spacetime. As a by-product, the spacetime with one singularity and one
singular sphere can also be modeled as a black hole with a firewall \cite{alm:2013}.

We shall construct our solutions in an especially attractive class of generalized Proca theories \cite{hei:2014, tas:2014}. The especial class is the nonlinear electrodynamics with gauge-invariant Lagrangian $L=K(F_{\mu\nu}F^{\mu\nu})$, where $K$ is an arbitrary function of $F_{\mu\nu}F^{\mu\nu}$. In the background of static and spherically symmetric spacetime, its energy-momentum tensor $T_{\mu\nu}$ has the symmetry of $T_{00}=T_{11}$ and thus constrains $g_{00}=1/g_{11}$ \cite{bro:2001}.
Such theories, in particular, the Born-Infeld
nonlinear electrodynamics, gained much attention as limiting cases
of certain models of string theory (see \cite{sei:1999, tse:1999} for reviews).
Depending on our circumstances we shall opt for the method of \cite{bea:1998}. It is the most efficient method for the construction of black hole solutions. Concretely,
one assume the metric initially and
then solve for the Maxwell field $\phi$ and the Lagrangian function $K(F_{\mu\nu}F^{\mu\nu})$.  It is
found the application of this method is quite a success and is widely used in the construction of the black hole solutions \cite{noj:2017,eli:2002,dym:2004,ans:2007,ans:2008,joh:2013,dym:2015,cul:2017,ma:2015,kun:2015,pra:2015,rod:2016,fan:2016,chi:2017,fer:2017,pon:2017,cai:2004}.

The paper is organized as follows. In section~\ref{sec:2}, we
derive the equations of motion in the theories of nonlinear electrodynamics. In
section~\ref{sec:3}, we construct the black hole spacetime with $N$ horizons. The spacetime is asymptotically flat in space. In section~\ref{sec:4}, we present the cosmos spacetime with $N$ horizons. In section~\ref{sec:5}, we study the black hole spacetime with $M$ horizons and $N$ singularities. Since the spacetime is static,  spherically symmetric and asymptotically flat, one could conceal all the singularities by the smallest horizon. So the cosmic censorship conjecture is not violated. In section~\ref{sec:6}, we give the solution of  multi-horizon black hole in multi-horizon universe. In section~\ref{sec:7}, we model a black hole spacetime with the firewall. Finally, conclusions and discussions are given in section~\ref{sec:8}. Throughout this paper, we adopt the
system of units in which $G=c=\hbar=1$ and the metric signature
$(-,\ +,\ +,\ +)$.
\section{equations of motion in nonlinear electrodynamics theories}
\label{sec:2}
We start from the action of nonlinear electrodynamics theories which are minimally coupled to gravity
\begin{equation}
S=\frac{1}{16\pi}\int\sqrt{-g}\left[R+K\left(X\right)\right]d^4x\;,
\end{equation}
with
\begin{equation}
X=F_{\mu\nu}F^{\mu\nu}\;,\ \ \ \  F_{\mu\nu}=\nabla_{\mu}A_{\nu}-\nabla_{\nu}A_{\mu}\;.
\end{equation}
Here $R$ is the Ricci scalar and $A_{\mu}$ is the Maxwell field. $K(X)$ is the function
of $X$ to be specified. Variation of the action with respect to the metric gives
the Einstein equations
\begin{eqnarray}
G_{\mu\nu}=-2K_{,X}F_{\mu\lambda}F_{\nu}^{\lambda}+\frac{1}{2}g_{\mu\nu}K\;, \ \ \ \ \ K_{,X}\equiv\frac{dK}{dX}\;.
\end{eqnarray}
Variation of the action with respect to the field $A_{\mu}$ gives the generalized
Maxwell equations
\begin{eqnarray}
\nabla_{\mu}\left(K_{,X}F^{\mu\nu}\right)=0\;.
\end{eqnarray}
In the background of static and spherically symmetric spacetime which can
always be parameterized as
\begin{eqnarray}
ds^2=-U\left(r\right)dt^2+\frac{1}{U\left(r\right)}dr^2+f\left(r\right)^2d\Omega_{2}^{2}\;.
\end{eqnarray}
Here $d\Omega_2^2= d\theta^2+\sin^2\theta d\phi^2$. The non-vanishing component of Maxwell field $A_{\mu}$ is uniquely to be
\begin{eqnarray}
A_0=\phi\left(r\right)\;,
\end{eqnarray}
by resorting to a gauge transformation of $A_{\mu}\rightarrow A_{\mu}+\nabla_{\mu}\psi$.
Then we obtain the Einstein equations and the generalized Maxwell equation
\begin{eqnarray}
-\frac{U^{'}f^{'}}{f}-\frac{2Uf^{''}}{f}+\frac{1}{f^2}-\frac{Uf^{'2}}{f^2}&=&2K_{,X}\phi^{'2}+\frac{1}{2}K\;, \label{7}\\
-\frac{U^{'}f^{'}}{f}+\frac{1}{f^2}-\frac{Uf^{'2}}{f^2}&=&2K_{,X}\phi^{'2}+\frac{1}{2}K\;, \label{8}\\
\frac{U^{'}f^{'}}{f}+\frac{Uf^{''}}{f}+\frac{1}{2}U^{''}&=&-\frac{1}{2}K\;, \label{9}\\
\left(f^2K_{,X}\phi^{'}\right)^{'}&=&0\;.\label{10}
\end{eqnarray}
The prime denotes the derivative with respect to $r$. Eqs.~(7-9) comes from $G_0^0=\rho$, $G_1^1=p_r$ and $G_2^2=p_{\theta}$, respectively.

Due to the Bianchi identities, only three of the four equations are independent. One usually assume
the expression of $K(X)$ initially. Then they are left with three unknown
functions, $U$, $f$ and $\phi$ and the system of equations are closed. However, we
shall not follow this way in this paper because the equations of motion are
rather involved. Instead, we shall let $K(X)$ keep open and assume the expression of the metric component $U(r)$ initially. Then we
solve for the Maxwell field $\phi$ and the unknown function $K(X)$. The method is quite a success and is widely used in
the construction of the black hole solutions. So in the next
subsections, we shall seek for some interesting spacetimes by using this method.
Before the presentation of some interesting spacetimes, we observe the
difference of Eq. (7) and Eq. (8) which gives
\begin{eqnarray}
f^{''}=0\;.
\end{eqnarray}
Thus we obtain the physical solution for $f$,
\begin{eqnarray}
f=r\;.
\label{12}
\end{eqnarray}
Therefore, the static and spherically symmetric spacetime with the source of nonlinear electrodynamic field is
\begin{eqnarray}
ds^2=-U\left(r\right)dt^2+\frac{1}{U\left(r\right)}dr^2+r^2d\Omega_{2}^{2}\;.\label{5}
\end{eqnarray}
\section{Black holes with N horizons}\label{sec:3}
In section~\ref{sec:2}, the function $f=r$ is obtained. So we could assume that the
metric for static and spherically symmetric black holes with $N$ horizons can be
written as Eq.~(\ref{5}) with
\begin{eqnarray}
U\left(r\right)&=&\left(1-\frac{a_1}{r}\right)\left(1-\frac{a_2}{r}\right)\left(1-\frac{a_3}{r}\right)
\cdot\cdot\cdot\left(1-\frac{a_N}{r}\right)\nonumber\\&=&\prod_i^{N}\left(1-\frac{a_i}{r}\right)\;.
\end{eqnarray}
Here $a_i$ are $N$ positive constants ($i = 1, 2, 3,\cdot\cdot\cdot, N$).
If $0 < a_1 < a_2 <\cdot\cdot\cdot < a_N$, there are $N$ horizons in the spacetime. The
spacetime is singular at $r = 0$ and asymptotically flat at $r=+\infty$.
Substituting the metric into the Einstein equations with the energy momentum tensor of anisotropic fluid,
\begin{eqnarray}
G_{\mu\nu}=8\pi T_{\mu\nu}\;,
\end{eqnarray}
we find the energy density $\rho$ and the pressures, $p_r, p_{\theta}, p_{\phi}$ of the fluid
\begin{eqnarray}
\rho&=&\frac{1}{8\pi}\left[\frac{1}{r^4}\sum_{i\neq j}^{N}a_ia_j-\frac{2}{r^5}\sum_{i\neq j\neq k}^{N}a_ia_ja_k\right.\nonumber\\&& \left.+\frac{3}{r^6}\sum_{i\neq j\neq k\neq l}^{N}a_ia_ja_ka_l\right.\nonumber\\&& \left.-\cdot\cdot\cdot+\frac{\left(-1\right)^N}{r^{N+2}}\left(N-1\right)a_1a_2a_3\cdot\cdot\cdot a_N\right]\;,\\
p_r&=&-\rho\;,\\
p_{\theta}&=&\frac{1}{8\pi}\left[\frac{1}{r^4}\sum_{i\neq j}^{N}a_ia_j-\frac{3}{r^5}\sum_{i\neq j\neq k}^{N}a_ia_ja_k\right.\nonumber\\&& \left.+\frac{6}{r^6}\sum_{i\neq j\neq k\neq l}^{N}a_ia_ja_ka_l-\cdot\cdot\cdot\right.\nonumber\\&& \left.+\frac{\left(-1\right)^N}{r^{N+2}}\frac{N\left(N-1\right)}{2}a_1a_2a_3\cdot\cdot\cdot a_N\right]\;,\\
p_{\phi}&=&p_{\theta}\;.
\end{eqnarray}
If $N$ is odd, the energy density would be negative for sufficiently small $r$. By contrast, if $N$ is even, the energy density is always positive for both large and small $r$.
The positive energy theorem claims that the energy density can
not be negative. Therefore we shall be interested in the even number of $N$
in the next subsection. As an example, we focus on the $N=4$ case.
\subsection{Black holes with 4 horizons}
The metric for static and spherically symmetric black holes with $4$ horizons
assumes the form of Eq.~(\ref{5}) with
\begin{eqnarray}
U=\left(1-\frac{a_1}{r}\right)\left(1-\frac{a_2}{r}\right)\left(1-\frac{a_3}{r}\right)\left(1-\frac{a_4}{r}\right)\;. \label{19}
\end{eqnarray}
It is asymptotically flat in space. We assume $0 < a_1 < a_2 < a_3 < a_4$. There are four horizons, $r = a_i$ and a
singularity, $r = 0$ in the spacetime.
The energy density $\rho$ and the pressures, $p_r, p_{\theta}, p_{\phi}$ of the anisotropic fluid
have the form
\begin{eqnarray}
\rho&=&\frac{1}{8\pi}\left(\frac{\alpha}{r^4}-\frac{2\beta}{r^5}+\frac{3\gamma}{r^6}\right)\;,\\
p_r&=&-\rho\;,\\
p_{\theta}&=&\frac{1}{8\pi}\left(\frac{\alpha}{r^4}-\frac{3\beta}{r^5}+\frac{6\gamma}{r^6}\right)\;,\\
p_{\phi}&=&p_{\theta}\;,
\end{eqnarray}
with
\begin{eqnarray}
\alpha&=&a_1a_2+a_1a_3+a_1a_4+a_2a_3+a_2a_4+a_3a_4\;,\\
\beta&=&a_1a_2a_3+a_1a_2a_4+a_1a_3a_4+a_2a_3a_4\;,\\
\gamma&=&a_1a_2a_3a_4\;.
\end{eqnarray}
The energy density is always positive provided that
\begin{eqnarray}
\beta^2-3\alpha\gamma<0\;.\label{delta}
\end{eqnarray}
This requirement is not hard to be satisfied. Substituting Eq.~(\ref{19}) into Eq.~(\ref{7}-\ref{10}), we obtain the corresponding electric potential $\phi$,  the square of field strength $F_{\mu\nu}F^{\mu\nu}$ and the Lagrangian function $K$:
\begin{eqnarray}
\phi&=&\phi_0\left(\frac{2\alpha}{r}-\frac{5\beta}{2r^2}+\frac{3\gamma}{r^3}\right)\;, \label{29}\\
F_{\mu\nu}F^{\mu\nu}&=&-2\phi_0^2\left(\frac{2\alpha}{r^2}-\frac{5\beta}{r^3}+\frac{9\gamma}{r^4}\right)^2\;,\label{30a}\\
K\left(F_{\mu\nu}F^{\mu\nu}\right)&=&-\frac{2\alpha}{r^4}+\frac{6\beta}{r^5}-\frac{12\gamma}{r^6}\;.\label{30b}
\end{eqnarray}
Here $\phi_0$ is an integration constant. Actually, $\phi_0$ is related to the electric
charge of the spacetime as follows
\begin{eqnarray}
Q_e\equiv r^2K_{,X}F^{01}=\frac{1}{2\phi_0}\;.
\end{eqnarray}
When
\begin{eqnarray}
&&a_1=m-\sqrt{m^2-Q^2}\;, \ \ \ \ \  a_3=a_4=0\;,\\&&
a_2=m+\sqrt{m^2-Q^2}\;,\ \ \ \ \ \ \phi_0=\frac{1}{2Q}\;,
\end{eqnarray}
we have
\begin{eqnarray}
Q_e=Q\;,\phi=\frac{Q}{r}\;,F_{\mu\nu}F^{\mu\nu}=-\frac{2Q^2}{r^4}\;,K=F_{\mu\nu}F^{\mu\nu}\;.
\end{eqnarray}
It is exactly the Maxwell theory for the Reissner-Nordstr$\ddot{\textrm{o}}$m spacetime. $m$, $Q$ are the mass and the electric charge of the black hole, respectively. Eq.~(\ref{29}) tells us the electric potential $\phi$ is endowed with two extra terms, $\beta/r^2,\ \gamma/r^3$ except for the usual Maxwell term, $\phi=\alpha/r$. With the presence of extra terms, the electric force exerted on a charged particle with unit electric charge becomes
\begin{eqnarray}
F_e=\phi_0\left(-\frac{2\alpha}{r^2}+\frac{5\beta}{r^3}-\frac{9\gamma}{r^4}\right)\;,
\end{eqnarray}
which differs from the Coulomb's law.

Eq.~(\ref{30a}) can be written as a biquadratic algebraic equation of $r$. By solving this equation we obtain $r=r(F^2)$. Then substituting this expression into Eq.~(\ref{30b}), we find $K$ can be explicitly expressed as the function of $F^2$.

\subsection{Penrose diagram}
In this subsection, we give the Penrose diagram of the 4-horizon spacetime.
One of the easy ways of constructing Penrose diagrams for any spherical (and not only spherical) space-times can be found in the book
of Bronnikov and Rubin \cite{bron:2012} (see also \cite{bron:1979}). But here we follow the ways of Plebanski and Krasinski \cite{ple:2006}

Following \cite{ple:2006}, we define the tortoise coordinate $r_{\ast}$ as follows
\begin{eqnarray}
r_{\ast}=r+\sum_i\frac{a_i^4\ln| r-a_i|}{\left(a_i-a_j\right)\left(a_i-a_k\right)\left(a_i-a_l\right)}\;.\label{totorse}
\end{eqnarray}
Here $i$ runs over from $1$ to $4$ and $i\neq j\neq k\neq l$. We make coordinates transformation $(t, r)\rightarrow (v, u)$
\begin{eqnarray}
v=e^{\gamma r_{\ast}}\sinh{\gamma t}\;,\ \ \ \ \ \
u=e^{\gamma r_{\ast}}\cosh{\gamma t}\;,\label{36a}
\end{eqnarray}
where $\gamma$ is a constant. Then the metric Eq.~(\ref{5}) becomes
\begin{eqnarray}
ds^2=F\left(v,u\right)\left(-dv^2+du^2\right)+r^2d\Omega_2^2\;,
\end{eqnarray}
with
\begin{eqnarray}
F=\frac{1}{\gamma r^4e^{2\gamma r}}\prod_i |r-r_i |^{1-\frac{2\gamma a_i^4}{\left(a_i-a_j\right)\left(a_i-a_k\right)\left(a_i-a_l\right)}}\;.
\end{eqnarray}
Here $i$ runs over from $1$ to $4$ and $i\neq j\neq k\neq l$. In order
to remove the coordinate singularity $a_i$, we should let
\begin{eqnarray}
{1-\frac{2\gamma_i a_i^4}{\left(a_i-a_j\right)\left(a_i-a_k\right)\left(a_i-a_l\right)}}=0\;,
\end{eqnarray}
namely,
\begin{eqnarray}
\gamma_i=\frac{1}{2a_i^4}{\left(a_i-a_j\right)\left(a_i-a_k\right)\left(a_i-a_l\right)}\;.
\end{eqnarray}
From Eq.~(\ref{36a}), we obtain
\begin{eqnarray}
u^2-v^2&=&e^{2\gamma r_{\ast}}\nonumber\\&=&\frac{e^{2\gamma r}\prod_i\left(r-r_i\right)}{\prod_i |r-r_i |^{1-\frac{2\gamma a_i^4}{\left(a_i-a_j\right)\left(a_i-a_k\right)\left(a_i-a_l\right)}}}\;.
\end{eqnarray}
Substituting $r=a_i$ and $\gamma=\gamma_i$ into above equation, we have
\begin{eqnarray}
u^2-v^2=0\;.
\end{eqnarray}
So in the $(v, u)$ coordinate system, the horizon $r =a_i$ is not
singular. Furthermore, the horizon consists of two lines $u=
\pm v$ in the $(v, u)$ plane. At the true singularity $r=0$, we have
\begin{eqnarray}
u^2-v^2=\textrm{constant}>0\;.
\end{eqnarray}
Therefore in the $(v, u)$ coordinates, they are a pair of hyperbolae and the hyperbolae are timelike. After employing the Penrose transformation, we arrive at the conventional coordinates system by
\begin{eqnarray}
\overline{V}&=&\frac{1}{2}\left[\tanh\left(u+v\right)-\tanh\left(u-v\right)\right]\;,\\
\overline{U}&=&\frac{1}{2}\left[\tanh\left(u+v\right)+\tanh\left(u-v\right)\right]\;.
\end{eqnarray}
Figure 1 shows the Penrose diagram of the black hole spacetime with four
horizons. The singularities are timelike.
\begin{figure}[h]
\begin{center}
\includegraphics[width=9cm]{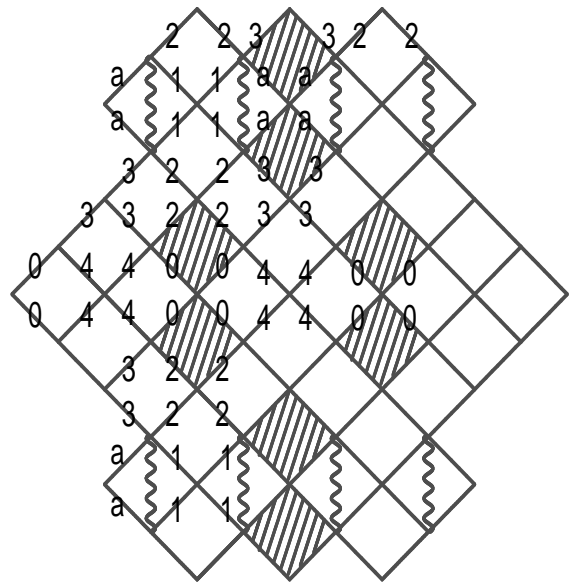}
\caption{The  Penrose diagram of a black hole with four
horizons. The wavy lines are the singularities. $1,2,3,4$ stand for the number of horizons, $0$ stands for the null spatial infinity and $a$ stands for $r=-\infty$. }\label{fig:1}
\end{center}
\end{figure}

\subsection{Observational aspect 1: quasinormal modes}

When a black hole is perturbed it evolves into three stages by emitting gravitational waves: 1) a relatively short period of initial outburst of radiation,
2) a long period of damping proper oscillations,
dominated by the so-called quasinormal
modes, 3) at very large time the quasinormal modes are suppressed
by power-law or exponential late-time tails. The dominating contribution
to gravitational waves is the quasinormal mode with
lowest frequency: the fundamental mode. So in this subsection we study the quasinormal modes generated by
the propagation of a test
minimally coupled massless scalar field in the 4-horizon black hole spacetime. To this end, we start from the well-known Klein-Gordon equation
\begin{eqnarray}
\nabla^2\Psi=0\;,
\end{eqnarray}
which is the general perturbation equation for the massless scalar field in the curve spacetime.
Here $\nabla^2$ is the four dimensional Laplace operator and $\Psi$ the massless scalar field. Making the standard decomposition
\begin{eqnarray}
\Psi=e^{-i\omega t}Y_{lm}\left(\theta,\ \phi\right)\frac{\Phi\left(r\right)}{r}\;,
\end{eqnarray}
we obtain the radial perturbation equation
\begin{eqnarray}
\frac{d^2\Phi}{d r_{\ast}^2}+\left(\omega^2-V\right)=0\;,
\end{eqnarray}
where the effective potential is given by
\begin{eqnarray}
V=U\left(\frac{l\left(l+1\right)}{r^2}+\frac{U_{,r}}{r}\right)\;,
\end{eqnarray}
and $r_{\ast}$ is the totorse coordinate defined by
\begin{eqnarray}
r_{\ast}=r+\sum_i\frac{a_i^4\ln {\left(r-a_i\right)}}{\left(a_i-a_j\right)\left(a_i-a_k\right)\left(a_i-a_l\right)}\;.
\end{eqnarray}
\begin{figure}[h]
\begin{center}
\includegraphics[width=9cm]{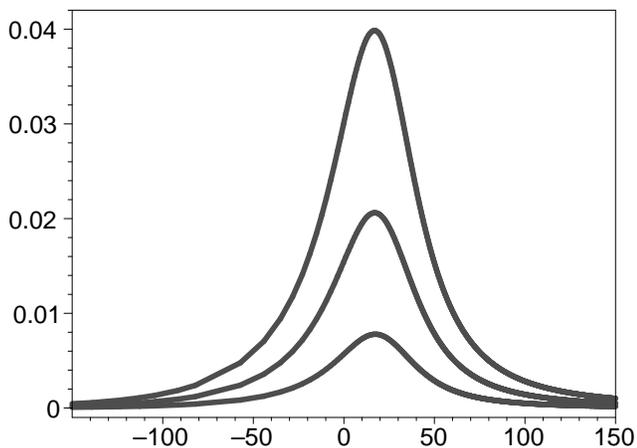}
\caption{The effective potential $V(r_{\ast})$ as a function of the tortoise coordinate $r_\ast$ assuming $a_1=1, a_2=2, a_3=3, a_4=4$ for three different cases $l=3,2,1$, from up to down, respectively.}\label{pot}
\end{center}
\end{figure}
The effective potential $V$ as a function of the tortoise coordinate $r_\ast$ can be seen in Fig.\ref{pot}.

We shall evaluate the quasinormal frequencies for the massless scalar field by
using the third-order WKB approximation, a numerical and perhaps the most popular method, devised by Schutz, Will and Iyer \cite{will:1985,will:1987,iyer:1987}.
This method has been used extensively in evaluating quasinormal frequencies of various black holes. For an incomplete list see \cite {quasi:99} and references therein.

The quasinormal frequencies are given by
\begin{eqnarray}
\omega^2=V_0+\Lambda\sqrt{-2V_0^{''}}-i\nu\left(1+\Omega\right)\sqrt{-2V_0^{''}}\;,
\end{eqnarray}
where $\Lambda$ and $\Omega$ are

\begin{eqnarray}
\Lambda&=&\frac{1}{\sqrt{-2V_0^{''}}}\left\{\frac{V_0^{(4)}}{V_0^{''}}\left(\frac{1}{32}+\frac{1}{8}\nu^2\right)
\right.\nonumber\\&&\left.-\left(\frac{V_0^{'''}}{V_0^{''}}\right)^2\left(\frac{7}{288}+\frac{5}{24}\nu^2\right)\right\}\;,\\
\Omega&=&\frac{1}{\sqrt{-2V_0^{''}}}\left\{\frac{5}{6912}\left(\frac{V_0^{'''}}{V_0^{''}}\right)^4\left(77+188\nu^2\right)
\right.\nonumber\\&&\left.-\frac{1}{384}\left(\frac{V_0^{'''2}V_0^{(4)}}{V_0^{''3}}\right)\left(51+100\nu^2\right)
\right.\nonumber\\&&\left.+\frac{1}{2304}\left(\frac{V_0^{(4)}}{V_0^{''}}\right)^2\left(67+68\nu^2\right)\right.\nonumber\\&&\left.
+\frac{1}{288}\left(\frac{V_0^{'''}V_0^{(5)}}{V_0^{''2}}\right)\left(19+28\nu^2\right)\right.\nonumber\\&&\left.-\frac{1}{288}\left(\frac{V_0^{(6)}}{V_0^{''}}
\left(5+4\nu^2\right)\right)\right\}\;,
\end{eqnarray}
and

\begin{eqnarray}
\nu=n+\frac{1}{2}\;,\ \ \ \ V_0^{(s)}=\frac{d^sV}{dr_{\ast}^s}|_{r_{\ast}=r_p}\;,
\end{eqnarray}
$n$ is overtone number and $r_p$ corresponds to the peak of the effective
potential. It is pointed that \cite{car:04} that the accuracy of the WKB method depends on the multipole
number $l$ and the overtone number $n$. The WKB approach is consistent with the numerical method very well proviede that  $l>n$. Therefore we shall
present the quasinormal frequencies of scalar perturbation for $n=0$ and $l=1,2,3$, respectively. In order to satisfy the constraint equation Eq.~(\ref{delta}), we set $a_1=1, a_2=2, a_3=3, 3\leq a_4\leq 4$.
The fundamental quasinormal frequencies of the massless scalar perturbation field for fixed $l=1,2,3$ are given by in table I. From the table we see that with the decreasing
of the outermost horizon, both the real part and the imaginary part of the frequencies are increasing. When $a_4=a_3=3$,
we obtain an extreme black hole.

\subsection{Observational aspect 2: black hole shadow}
As another potential observational aspect, we consider the black hole shadow in this subsection. When a black hole is in front of a bright background which is produced by faraway radiating object, it casts a black  shadow. The apparent shape of the black hole is just defined by the boundary of the black
shadow. The ability of very long baseline interferometry (VLBI) observation has been improved significantly. This opens up the possibility to make observations on the black hole shadow \cite{bar:73, fal:00}. A significant attention has been payed to the study of black
hole shadow and it has become a quite active research field \cite{shadow} (for a review, see \cite{boz:10} ).

\begin{figure}[h]
\begin{center}
\includegraphics[width=9cm]{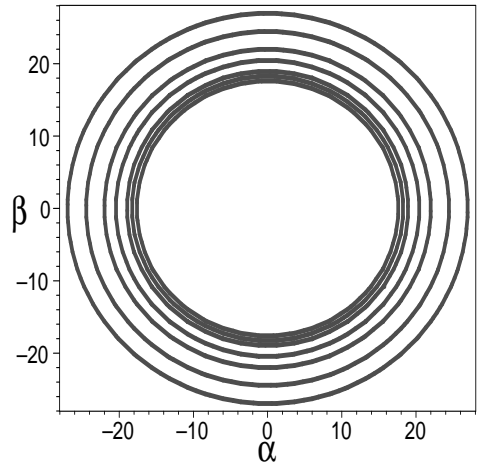}
\caption{Shadow of a 4-horizon black hole for different radii, from inner to outer. $(1)$. $a_4=4,a_3=3,a_2=2,a_1=1$; $(2)$. $a_4=4,a_3=3,a_2=2,a_1=1.5$ ;
$(3)$. $a_4=4,a_3=3,a_2=2,a_1=2$; $(4)$. $a_4=4,a_3=3,a_2=2.5,a_1=2.5$; $(5)$. $a_4=4,a_3=3,a_2=3,a_1=3$; $(6)$. $a_4=4,a_3=3.5,a_2=3.5,a_1=3.5$; $(7)$. $a_4=4,a_3=4,a_2=4,a_1=4$. }\label{shadow}
\end{center}
\end{figure}

\begin{figure}[h]
\begin{center}
\includegraphics[width=9cm]{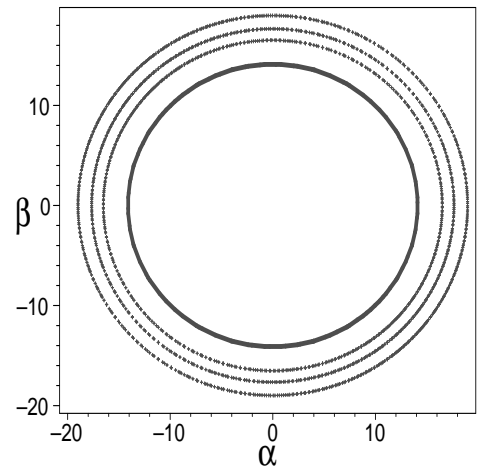}
\caption{Shadow of a Reissner-Nordstrom  (solid line) and 4-horizon black hole (dotted lines) for different radii, from inner to outer. $(1)$. $a_4=4,a_3=3,a_2=0,a_1=0$; $(2)$. $a_4=4,a_3=3,a_2=2,a_1=0$ ;
$(3)$. $a_4=4,a_3=3,a_2=2,a_1=1$; $(4)$. $a_4=4,a_3=3,a_2=2,a_1=2$.}\label{shadowRN}
\end{center}
\end{figure}

In order to obtain the boundary of the shadow of the black hole, we need to study the radial motion
\begin{eqnarray}
\dot{r}^2+V_{eff}=0\;,
\end{eqnarray}
where the dot denotes the derivative respect to the proper time and the effective potential is
\begin{eqnarray}
V_{eff}=\frac{1}{r^2}U\left(\eta+\xi^2\right)E^2-E^2\;.
\end{eqnarray}
Here $\mathscr{K}$ is the Carter constant \cite{carter:68}. $E$ and $L$ are energy and angular momentum of the photon, respectively.
$\eta=\mathscr{K}/E^2$ and $\xi=L/E$ are two impact parameters which character the motion of photon near the black hole.
The boundary of the shadow is mainly determined by the
circular photon orbit, which satisfies the following conditions
\begin{eqnarray}
V_{eff}=0\;,\ \ \ \ \ \ \ \partial_rV_{eff}=0.
\end{eqnarray}
In order to find the shadow of the 4-dimensional black hole, we introduce the celestial coordinates
\begin{eqnarray}
\alpha=\lim_{r\rightarrow\infty}\left(\frac{rP^{(\phi)}}{P^{(t)}}\right)\;,\ \ \ \ \beta=\lim_{r\rightarrow\infty}\left(\frac{rP^{(\theta)}}{P^{(t)}}\right)\;,
\end{eqnarray}
where $(P^{(t)}, P^{(\theta)}, P^{(\phi)})$ are the vi-tetrad components of angular momentum. Supposing the observer is located in the equatorial plane of the black hole, one has $\theta=\pi/2$. Then the celestial coordinates $\alpha$ and $\beta$ takes the form

\begin{eqnarray}
\alpha=-\xi\;,\ \ \ \ \beta=\pm\sqrt{\eta}\;.
\end{eqnarray}
Let the parameters ($\xi,\ \eta$) run over all possible values. Then we obtain the shadow of black hole in the parameter space ($\alpha, \beta$).
In Fig.~\ref{shadow}, we plot the shadow of a 4-dimensional black hole for different radii. The figure
shows that the size of the shadow increases with the increasing of the inner horizons. In Fig.~\ref{shadowRN}, we plot the shadow of a Reissner-Nordstrom  (solid line) and a 4-dimensional black hole (dotted lines) for different radii. It shows that the size of the shadow becomes greater due to the presence of inner horizons. We hope this difference may be confirmed by the future VLBI observations.

\section{Cosmos with N horizons}\label{sec:4}
In previous section, we present the black hole spacetime with multiple horizons. In this section, we turn to the case of cosmos. We know both the de Sitter universe and the Friedmann-Robertson-Walker Universe have one apparent horizon. Then can we construct a cosmos with multiple horizons? We focus on this question in this section. We assume the metric for static and spherically symmetric spacetimes with $N$ cosmic horizons could be written as Eq.~(\ref{5}) with
\begin{eqnarray}
U\left(r\right)&=&\left(1-h_1r\right)
\left(1-h_2r\right)\left(1-h_3r\right)\cdot\cdot\cdot\left(1-h_Nr\right)\nonumber\\&=&\prod_i^N\left(1-h_i r\right)\;.
\end{eqnarray}
Here $h_i$ are $N$ positive constants ($i = 1, 2, 3,\cdots,N$). If $0 < h_1 < h_2 < \cdot\cdot\cdot<
h_N$, there are $N$ cosmic horizons. When $h_i=0$, the metric reduces to the Minkowsky one.
The energy density $\rho$ and the pressures, $p_r, p_{\theta}, p_{\phi}$ of the source are given by
\begin{eqnarray}
\rho&=&\frac{1}{8\pi}\left[\frac{2}{r}\sum_{i}^{N}h_i-3\sum_{i\neq j}^{N}h_ih_j+{4r}\sum_{i\neq j\neq k}^{N}h_ih_jh_k-\cdot\cdot\cdot\right.\nonumber\\&& \left.+{\left(-1\right)^{N-1}}\left(N+1\right){r^{N-2}}h_1h_2h_3\cdot\cdot\cdot h_N\right]\;,\\
p_r&=&-\rho\;,\\
p_{\theta}&=&\frac{1}{8\pi}\left[-\frac{1}{r}\sum_{i}^{N}h_i+3\sum_{i\neq j}^{N}h_ih_j-{6r}\sum_{i\neq j\neq k}^{N}h_ih_jh_k-\cdot\cdot\cdot\right.\nonumber\\&& \left.+{\left(-1\right)^{N}}\frac{N\left(N+1\right)}{2}{r^{N-2}}h_1h_2h_3\cdot\cdot\cdot h_N\right]\;,\\\;
p_{\phi}&=&p_{\theta}\;.
\end{eqnarray}
It seems the spacetime is singular at $r=0$ because of the divergence of density at $r=0$ when
\begin{eqnarray}
\sum_i^Nh_i \neq 0\;.
\end{eqnarray}
This is not the case. We shall illustrate this point below by using an example with $N=3$.

If $N$ is even, the energy density can be negative for sufficiently large $r$. In
contrast, if $N$ is odd, the energy density is always positive at both large
and small $r$. Therefore, we are only interested in cases with $N$ being odd numbers. As an example, we shall only focus on the case with $N=3$ below.

\subsection{Cosmos with 3 horizons}
The metric for static and spherically symmetric spacetimes with $3$ cosmic
horizons could be written as Eq.~(\ref{5}) with
\begin{eqnarray}
U\left(r\right)=\left(1-h_1r\right)\left(1-h_2r\right)\left(1-h_3r\right)\;.\label{19a}
\end{eqnarray}
We assume $0<h_1<h_2<h_3$. There are three cosmic horizons in the spacetime.
The energy density $\rho$ and the pressures, $p_r, p_{\theta}, p_{\phi}$ of the anisotropic fluid
have the form
\begin{eqnarray}
\rho&=&\frac{1}{8\pi}\left(\frac{2\alpha}{r}-{3\beta}+{4\gamma r}\right)\;,\\
p_r&=&-\rho\;,\\
p_{\theta}&=&\frac{1}{8\pi}\left(-\frac{\alpha}{r}+{3\beta}-{6\gamma r}\right)\;,\\
p_{\phi}&=&p_{\theta}\;,
\end{eqnarray}
with
\begin{eqnarray}
\alpha&=&h_1+h_2+h_3\;,\\
\beta&=&h_1h_2+h_1h_3+h_2h_3\;,\\
\gamma&=&h_1h_2h_3\;.
\end{eqnarray}
The energy density is always positive provided that
\begin{eqnarray}
9\beta^2-32\alpha\gamma<0\;.
\end{eqnarray}
Substituting Eq.~(\ref{19a}) into Eq.~(\ref{7}-\ref{10}), we obtain the consistent solution
\begin{eqnarray}
\phi&=&\phi_0\left(-{\alpha}{r^2}+{\gamma}{r^4}\right)\;,\\
F_{\mu\nu}F^{\mu\nu}&=&-8\phi_0^2r^2\left({2\gamma}{r^2}-{\alpha}\right)^2\;,\label{78a}\\
K\left(F_{\mu\nu}F^{\mu\nu}\right)&=&\frac{2\alpha}{r}-{6\beta}+{12\gamma}{r}\;.\label{78b}
\end{eqnarray}
Here $\phi_0$ is an integration constant, and it is related to the electric
charge of the spacetime
\begin{eqnarray}
Q_e\equiv r^2K_{,X}F^{01}=\frac{1}{4\phi_0}\;.
\end{eqnarray}
If
\begin{eqnarray}
&&h_3=0\;,\ \ \ \ \ h_1=-\sqrt{\frac{\lambda}{3}}\;,\ \ \ \ \ h_2=\sqrt{\frac{\lambda}{3}}\;,
\end{eqnarray}
we have
\begin{eqnarray}
\phi=0\;,\ \ \ \ \ F_{\mu\nu}F^{\mu\nu}=0\;,\ \ \ \ \ K\left(F_{\mu\nu}F^{\mu\nu}\right)=-2\lambda\;.
\end{eqnarray}
It is exactly the de Sitter (or Anti-de Sitter) solution.

Eq.~(\ref{78a}) can be written as a cubic algebraic equation of $r$. Solving this equation for $r=r(F^2)$ and substituting the expression into Eq.~(\ref{78b}), we find $K$ can be also explicitly expressed as the function of $F^2$.

\subsection{Penrose diagram}
To plot the Penrose diagram, we introduce the tortoise coordinate $r_{\ast}$ as
\begin{eqnarray}
r_{\ast}=-\sum_i\frac{\ln| 1-h_i r|}{\left(h_i-h_j\right)\left(h_i-h_k\right)}\;.
\end{eqnarray}
Here $i$ runs over from $1$ to $3$ and $i\neq j\neq k$. We make coordinates transformation $(t, r)\rightarrow (v, u)$
\begin{eqnarray}
v=e^{\gamma r_{\ast}}\sinh{\gamma t}\;,\ \ \ \ \ \
u=e^{\gamma r_{\ast}}\cosh{\gamma t}\;.\label{36a}
\end{eqnarray}
Then the metric Eq.~(\ref{5}) becomes
\begin{eqnarray}
ds^2=F\left(v,u\right)\left(-dv^2+du^2\right)+r^2d\Omega_2^2\;,
\end{eqnarray}
with
\begin{eqnarray}
F=\frac{1}{\gamma^2}\prod_i |1-h_i r |^{1+\frac{2\gamma h_i}{\left(h_i-h_j\right)\left(h_i-h_k\right)}}\;.
\end{eqnarray}
Now we see the spacetime is regular at $r=0$.
In order to remove the coordinate singularity $r_i$, we should let
\begin{eqnarray}
{1+\frac{2\gamma_i h_i}{\left(h_i-h_j\right)\left(h_i-h_k\right)}}=0\;,
\end{eqnarray}
namely,
\begin{eqnarray}
\gamma_i=-\frac{1}{2h_i}{\left(h_i-h_j\right)\left(h_i-h_k\right)}\;.
\end{eqnarray}
From Eq.~(\ref{36a}), we obtain
\begin{eqnarray}
u^2-v^2&=&e^{2\gamma r_{\ast}}\nonumber\\&=&\frac{\gamma\prod_i{\left(1-h_i r\right)}}{{\prod_i |1-h_i r |^{1+\frac{2\gamma}{\left(h_i-h_j\right)\left(h_i-h_k\right)}}}}\;.
\end{eqnarray}
Substituting $r=1/h_i$ and $\gamma=\gamma_i$ into above equation, we have
\begin{eqnarray}
u^2-v^2=0\;.
\end{eqnarray}
So in the $(v, u)$ coordinate system, the horizon $r =1/h_i$ is not
a singularity. Furthermore, it consists of two lines $u=
\pm v$ in the $(v, u)$ plane.
Figure 5 shows the Penrose diagram of the cosmos with $3$ horizons.

\begin{figure}[h]
\begin{center}
\includegraphics[width=9cm]{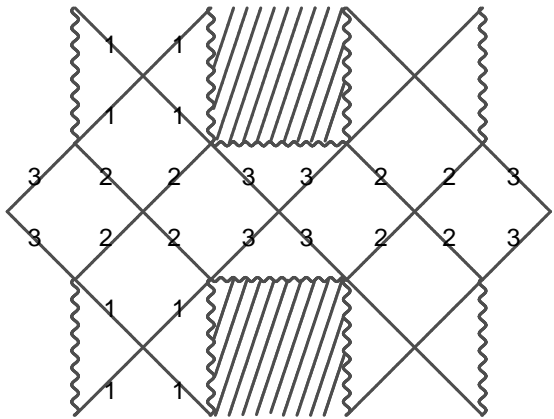}
\caption{The  Penrose diagram of a cosmos with three
horizons. The vertical wavy lines are the timelike infinities, $r=-\infty$.  The horizontal wavy lines are the spacelike infinities, $r=\infty$. $1,2,3$ stand for the number of horizons.}\label{fig:2a}
\end{center}
\end{figure}

\section{Black holes with $N$ singularities and $M$ horizons}\label{sec:5}
As argued in the introduction, the spacetime with multiple singularities and multiple horizons is highly likely to be produced in the merger process of multiple black holes. But it is rather involved to construct the analytic spacetime solutions describing this process. So far only a few solutions are present \cite{har:1972,kas:1993,bri:1994,gao:2004,shi:1999,mae:2009,mae:2010,gib:2010}.
In this section, we give the exact but toy-model spacetime with multiple singularities and multiple horizons.  As toy models, the solutions are all static and spherically symmetric. Therefore, the multiple horizons are concentric spheres and the singularities consists of one singular point and
multi-singular spheres. Some of these spacetimes may be constructed by using the Majumdar-Papapetrou solution \cite{har:1972}.

In Majumdar-Papapetrou solution, a system of extreme Reissner-Nordstrom black holes (ERN) (each having
electric charge equal to its mass) can remain in static equilibrium. No matter how the black holes are arranged in space, the electrostatic repulsions exactly balance
the gravitational attractions. In other words, the black holes in Majumdar-Papapetrou solution ignore each other. Now suppose one ERN black hole is in the center and
a lot of ERN black holes locate on a sphere as shown in Fig.6.  We expect a spacetime with a central singularity, a singular sphere (wavy circle) and three horizons (solid circles) is present.
\begin{figure}[h]
\begin{center}
\includegraphics[width=9cm]{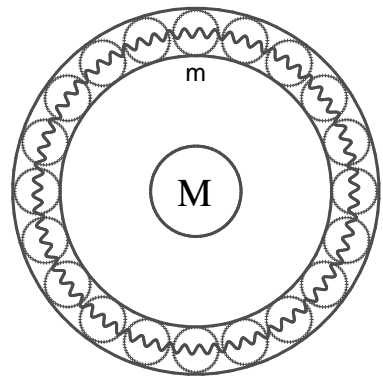}
\caption{A sketch of spacetime with a central singularity, a singular sphere (wavy circle) and three horizons (three solid circles). The spacetime is generated by a ERN in the center and a lot of ERNs on a sphere.}\label{fig:roll}
\end{center}
\end{figure}

The metric with multiple horizons and multiple singularities can be assumed to be Eq.~(\ref{5}) with
\begin{eqnarray}
U\left(r\right)&=&\frac{\left(1-\frac{b_1}{r}\right)\left(1-\frac{b_2}{r}\right)
\left(1-\frac{b_3}{r}\right)\cdot\cdot\cdot\left(1-\frac{b_{M}}{r}\right)}
{\left(1-\frac{a_1}{r}\right)\left(1-\frac{a_2}{r}\right
)\left(1-\frac{a_3}{r}\right)\cdot\cdot\cdot\left(1-\frac{a_{N-1}}{r}\right)}
\nonumber\\&=&\frac{\prod_i^{M}\left(1-\frac{b_i}{r}\right)}
{\prod_i^{N-1}\left(1-\frac{a_i}{r}\right)}\;.
\end{eqnarray}
We assume $0<a_1 < a_2 < a_3 <\cdot\cdot\cdot< a_{N-1} < b_1 < b_2 <\cdot\cdot\cdot< b_M$ and
$0< N-1<M$. Then there are $M$ black hole horizons, $r_i=b_i$ and $N$
singularities, $r_i=a_i$ together with $r=0$. All the singularities are concealed by the black hole horizons. We emphasize that the number $M$ and $N$ can not be arbitrary in order that the energy density is positive.  As for Fig.~6, the metric can be assumed to be Eq.~(\ref{5}) with
\begin{eqnarray}
U\left(r\right)&=&\frac{\left(1-\frac{b_1}{r}\right)^2\left(1-\frac{b_2}{r}\right)^2
\left(1-\frac{b_3}{r}\right)^2}
{\left(1-\frac{a_1}{r}\right)^2}\;.
\end{eqnarray}.

In principle, one could also construct black hole solutions with one or more singularities concealed by two horizons. For example, one could assume $b_1<a_{N-1} <b_2$ rather than $a_{N-2}<a_{N-1}<b_1$ in Eq. (77). In this case, the
singularity $a_{N-1}$ is concealed by two horizons $b_1$ and $b_2$.

\subsection{Black hole with 2 singularities and 2 horizons}
As an example, we focus on the case with $N=2$ and $M=2$ which is for the
Reissner-Nordstr$\ddot{\textrm{o}}$m black hole with a singular sphere inside the horizons.
Then the metric is given by
\begin{eqnarray}
ds^2&=&-\frac{\left(1-\frac{b_1}{r}\right)\left(1-\frac{b_2}{r}\right)}
{1-\frac{a_1}{r}}dt^2+\frac{1-\frac{a_1}{r}}{\left(1-\frac{b_1}{r}\right)
\left(1-\frac{b_2}{r}\right)}dr^2\nonumber\\&&+r^2d\Omega_2^2\;.\label{79}
\end{eqnarray}
When $a_1=0$, it reduces to the Reissner-Nordstr$\ddot{\textrm{o}}$m black hole. When $a_1\neq 0$,
a singular sphere is present inside the black hole.
The energy density $\rho$ and the pressures, $p_r, p_{\theta}, p_{\phi}$ of the matter source are
\begin{eqnarray}
\rho&=&\frac{1}{8\pi}\frac{\left(b_2-a_1\right)\left(b_1-a_1\right)}
{r^2\left(r-a_1\right)^2}\;,\\
p_r&=&-\rho\;,\\
p_{\theta}&=&-\frac{1}{8\pi}\frac{\left(b_2-a_1\right)\left(b_1-a_1\right)}
{r\left(r-a_1\right)^3}\;,\\
p_{\phi}&=&p_{\theta}\;.
\end{eqnarray}
It is apparent the energy density is always positive provided that $0<a_1<
b_1 < b_2$. When $b_2 = 0$ or $b_1 = 0$, the energy density is negative which
violates the positive energy theorem. On the other hand, when $b_1 = b_2 = 0$,
the singularity $r = a_1$ would be naked which violates the cosmic censorship
conjecture. Therefore, in order to obey both the positive energy theorem and
the cosmic censorship conjecture, there are at least two horizons to conceal
the two singularities. We note that both the energy density and pressures
are divergent at the singularity and the singular sphere.
The solution of Eq. (\ref{79}) corresponds to the generalized Maxwell theories
as follows
\begin{eqnarray}
\phi&=&\frac{\phi_0\left(3a_1-4r\right)}{\left(r-a_1\right)^2}\;,\\
F_{\mu\nu}F^{\mu\nu}&=&-\frac{8\phi_0^2\left(a_1-2r\right)^2}{\left(r-a_1\right)^6}\;,\label{98a}\\
K\left(F_{\mu\nu}F^{\mu\nu}\right)&=&-\frac{2\left(b_2-a_1\right)
\left(b_1-a_1\right)}{r\left(r-a_1\right)^3}\;,\label{98b}
\end{eqnarray}
where $\phi_0$ is an integration constant. The electric charge of the spacetime is
\begin{eqnarray}
Q_e\equiv r^2K_{,X}F^{01}=-\frac{\left(b_2-a_1\right)
\left(b_1-a_1\right)}{4\phi_0}\;.
\end{eqnarray}
When
\begin{eqnarray}
&&b_1=m-\sqrt{m^2-Q^2}\;, \ \ \ \ \ a_1=0 \;,\\&&
b_2=m+\sqrt{m^2-Q^2}\;,\ \ \ \ \ \phi_0=-\frac{Q}{4}\;,
\end{eqnarray}
they reduce to
\begin{eqnarray}
Q_e=Q\;,\ \ \  \ \ \ \ K\left(F_{\mu\nu}F^{\mu\nu}\right)=F_{\mu\nu}F^{\mu\nu}\;,
\end{eqnarray}
which is exactly the Maxwell theory for the Reissner-Nordstr$\ddot{\textrm{o}}$m spacetime.

Eq.~(\ref{98b}) can be written as a cubic algebraic equation of $r$. Solving this equation for $r=r(F^2)$ and substituting the expression into Eq.~(\ref{98a}), we find $F^2$ can be explicitly expressed as the function of $K$.

\subsection{Penrose diagram}
Define the tortoise coordinate $r_{\ast}$ as follows
\begin{eqnarray}
r_{\ast}&=&r+\frac{1}{b_1-b_2}\left[\left(a_1b_2-b_2^2\right)\ln|r-b_2|
\right.\nonumber\\&& \left.-\left(a_1b_1-b_1^2\right)\ln|r-b_1|\right])\;.
\end{eqnarray}
We make coordinates transformation $(t, r)\rightarrow (v, u)$
\begin{eqnarray}
v=e^{\gamma r_{\ast}}\sinh{\gamma t}\;,\ \ \ \ \ \
u=e^{\gamma r_{\ast}}\cosh{\gamma t}\;.\label{36b}
\end{eqnarray}
Then the metric Eq.~(\ref{5}) becomes
\begin{eqnarray}
ds^2=F\left(v,u\right)\left(-dv^2+du^2\right)+r^2d\Omega_2^2\;,
\end{eqnarray}
with
\begin{eqnarray}
F=\frac{1}{\gamma^2 r\left(r-a_1\right)e^{2\gamma r}}\prod_i |r-b_i|^{1-\frac{2\gamma b_i\left(b_i-a_1\right)}{b_i-b_j}}\;.
\end{eqnarray}
Here $i$ runs over from $1$ to $2$ and $i\neq j$. In order
to remove the coordinate singularity $r_i$, we should let
\begin{eqnarray}
1-\frac{2\gamma_i b_i\left(b_i-a_1\right)}{b_i-b_j}=0\;,
\end{eqnarray}
namely,
\begin{eqnarray}
\gamma_i=\frac{b_i-b_j}{2\gamma_i b_i\left(b_i-a_1\right)}\;.
\end{eqnarray}
From Eq.~(\ref{36b}), we obtain
\begin{eqnarray}
u^2-v^2&=&e^{2\gamma r_{\ast}}\nonumber\\&=&\frac{e^{2\gamma r}\prod_i\left(r-b_i\right)}{\prod_i |r-b_i|^{1-\frac{2\gamma b_i\left(b_i-a_1\right)}{b_i-b_j}}}\;.
\end{eqnarray}
Substituting $r=b_i$ and $\gamma=\gamma_i$ into above equation, we have
\begin{eqnarray}
u^2-v^2=0\;.
\end{eqnarray}
So in the $(v, u)$ coordinate system, the horizon $r =b_i$ is not
nonsingular. It consists of two lines $u=
\pm v$ in the $(v, u)$ plane. At the true singularity $r=a_1$ and $r=0$, we have
\begin{eqnarray}
u^2-v^2=\textrm{constant}>0\;,\ \ \  u^2-v^2=\textrm{constant}<0\;,
\end{eqnarray}
respectively.
Therefore in the $(v, u)$ coordinates, they are a pair of hyperbolae and the hyperbolae are timelike and spacelike, respectively.
Figure 3 shows the Penrose diagram of the black hole spacetime with two
horizons.

\begin{figure}[h]
\begin{center}
\includegraphics[width=9cm]{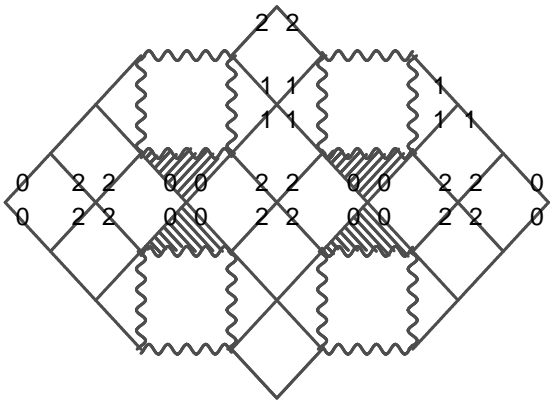}
\caption{The Penrose diagram for a black hole with two
singularities and two horizons. The vertical and horizontal wavy lines are the singularities $r=a_1$ and $r=0$, respectively. $1$, $2$ stand for  the number of horizons and $0$ stands for the null spatial infinity.}\label{fig:3}
\end{center}
\end{figure}

\section{Multi-horizon black hole in multi-horizon universe }\label{sec:6}

The black hole spacetime with multiple singularities and multiple horizons given in
the above section is asymptotically flat. In this section, we show that is possible to construct a
black hole spacetime with multiple black hole horizons and multiple cosmic horizons, which is not asymptotically flat.
A metric for multi-horizon black hole in multi-horizon universe can be given by Eq.~(\ref{5}) with
\begin{eqnarray}
U\left(r\right)&=&\left(1-\frac{a_1}{r}\right)\cdot\left(1-\frac{a_2}{r}\right)\cdot
\cdot\cdot\left(1-\frac{a_N}{r}\right)\nonumber\\&&\cdot\left(1-\frac{r}{b_1}\right)\cdot\left(1-\frac{r}{b_2}\right)
\cdot\cdot\cdot\left(1-\frac{r}{b_M}\right)\nonumber\\&=&\prod_i^{N}\left(1-\frac{a_i}{r}\right)\prod_j^{M}\left(1-\frac{r}{b_j}\right)\;.
\end{eqnarray}
Here we assume $0 < a_1 < a_2 < a_3 <\cdots< a_{N} < b_1 < b_2 < b_3 <\cdots< b_{M}<\infty$.
Then the spacetime has $M$ cosmic horizon $r_i=b_{i}$, $N$ black hole horizons $r_i=a_i$ and a singularity $r=0$. The reconstruction of its source in the framework of nonlinear electromagnetic field theories can be done
following the procedures shown in previous sections. Here we do not go into the detailed calculations.

\section{Black hole firewall}\label{sec:7}

In this section, we construct a black hole spacetime with the so-called firewall by using the nonlinear electromagnetic field theories. In order to resolve the black hole information loss paradox \cite{haw:1976}, Almheiri,
Marolf, Polchinski and Sully (hereinafter, AMPS) raised the black hole firewall proposal \cite{alm:2013}. In essence, AMPS argued that the local quantum field
theory, unitary, and no-drama (the infalling observers could not experience
anything unusual when crossing the event horizon) cannot all be consistent
with each other. They found that the most conservative resolution to this inherent inconsistency is to give up no-drama. Instead, there is a firewall with
considerable high energy density on or near the event horizon. As a result,
an infalling observer would be terminated once he/she hits the firewall.
Below we construct the model for a black hole with firewall by using
the nonlinear electromagnetic field theories. So let's start from Eq.~(\ref{5}).

\subsection{Thin wall}
The massive shell scenario of black hole firewall has been examined by a
number of authors (e.g., \cite{shell:1,shell:2,shell:3,shell:4}), the calculations are a standard application of Einstein¡¯s equations. In this subsection,
we assume the firewall is a shell of matter positioned at $r=r_0$ and the thickness of the shell is zero. We note that an exact solution describing a black hole surrounded by a thin massive shell has been investigated by Frauendiener et al. \cite{shell:5}.

Then the expression of density for the wall (or shell)
is given by the Dirac function
\begin{eqnarray}
\rho=\frac{m_1}{4\pi r^2}\delta\left(r-r_0\right)\;,\label{101}
\end{eqnarray}
such that the energy of the wall is
\begin{eqnarray}
\int_0^{\infty}4\pi r^2\rho dr=m_1\;.
\end{eqnarray}
Here the background metric Eq.~(\ref{5}) is taken into considerations. Substituting Eq.(\ref{101}) into the Einstein equations, we obtain
\begin{eqnarray}
U=1-\frac{2m_1}{r}\textrm{Heaviside}\left(r-r_0\right)-\frac{2m_2}{r}\;,
\end{eqnarray}
where the Heaviside function is related to the Dirac function by
\begin{eqnarray}
\left[\textrm{Heaviside}\left(r-r_0\right)\right]^{'}
=\delta\left(r-r_0\right)\;,
\end{eqnarray}
and can be expressed as
\begin{equation}\label{clipstr}
\textrm{Heaviside}\left(r-r_0\right)=\left\{\def\arraystretch{1.2}
  \begin{array}{@{}c@{\quad}l@{}}
   1\;,  & \text{if $ r>r_0$ }\;;\\
   undefined\;, & \text{if $ r=r_0$}\;;\\
   0\;, & \text{if $ r<r_0$ }\;.\\
  \end{array}\right.
\end{equation}
Here $m_2$ is the mass of the black hole and $m_1$ is the mass (or energy) of
the firewall. The metric is then given by
\begin{eqnarray}
ds^2&=&-\left[1-\frac{2m_1}{r}\textrm{Heaviside}\left(r-r_0\right)-\frac{2m_2}{r}\right]dt^2\nonumber\\&&
+\left[1-\frac{2m_1}{r}\textrm{Heaviside}\left(r-r_0\right)-\frac{2m_2}{r}\right]^{-1}dr^2\nonumber\\&&+r^2d\Omega^2\;.
\end{eqnarray}
It is the same as the solution in \cite{shell:5}.  The firewall is supposed in the vicinity of event horizon which corresponds to $r_0\geq 2m_2$. We show in the next section that the energy $m_1$ of the firewall could come from the electric field.

\subsection{Thick wall}
The Dirac function can be expressed as
\begin{equation}
\delta\left(r\right)=\lim_{\epsilon\rightarrow 0}\frac{1}{\pi}\frac{\epsilon}{r^2+\epsilon^2}\;,
\end{equation}
where $\epsilon$ is a constant.
Then the density of the wall can be assumed as
\begin{equation}
\rho=\frac{m_1}{4\pi r^2}\frac{1}{\pi}\frac{\epsilon}{\left(r-r_0\right)^2+\epsilon^2}\;.\label{85a}
\end{equation}
Substituting Eq. (\ref{85a}) into the Einstein equations, we obtain
\begin{equation}
U=1-\frac{2m_1}{r}\left[\frac{1}{2}+\frac{1}{\pi}{\arctan\left(\frac{r}{\epsilon}-\frac{r_0}{\epsilon}
\right)}\right]-\frac{2m_2}{r}\;.
\end{equation}
The metric is then given by
\begin{eqnarray}
&&ds^2=-\left\{1-\frac{2m_1}{r}\left[\frac{1}{2}+\frac{1}{\pi}{\arctan\left(\frac{r}{\epsilon}-\frac{r_0}{\epsilon}
\right)}\right]-\frac{2m_2}{r}\right\}dt^2\nonumber\\&&
+\left\{1-\frac{2m_1}{r}\left[\frac{1}{2}+\frac{1}{\pi}{\arctan\left(\frac{r}{\epsilon}-\frac{r_0}{\epsilon}
\right)}\right]-\frac{2m_2}{r}\right\}^{-1}dr^2\nonumber\\&&+r^2d\Omega^2\;.
\end{eqnarray}
Here $m_2$ is the mass of the black hole and $m_1$ is the mass (or energy) of
the firewall because  $\int_0^{\infty}4\pi r^2\rho dr=m_1$. The solution corresponds to the generalized Maxwell theory as follows:
\begin{eqnarray}
&&\phi=\frac{1}{2\pi}\frac{\phi_0 r \epsilon}{\left(r-r_0\right)^2+\epsilon^2}
\nonumber\\&&-\frac{3\phi_0}{2\pi}\arctan\left(r\epsilon^{-1}-r_0\epsilon^{-1}\right)\;,\\
&&F_{\mu\nu}F^{\mu\nu}=-\frac{2\phi_0^2\epsilon^2\left(2r^2-3rr_0+r_0^2
+\epsilon^2\right)^2}{\pi^2\left[\left(r-r_0\right)^2+\epsilon^2\right]^4}\;,\label{125a}\\
&&K\left(F_{\mu\nu}F^{\mu\nu}\right)=-\frac{4m_1\epsilon\left(r-r_0\right)}{\pi r\left[\left(r-r_0\right)^2+\epsilon^2\right]^2}\;,\label{125b}
\end{eqnarray}
The electric charge on the wall is
\begin{eqnarray}
Q_e=r^2K_{,X}F^{01}=\frac{m_1}{\phi_0}\;.
\end{eqnarray}
It is proportional to the total energy, $m_1$ of the wall. Thus we can conclude that the energy of the firewall could come from the electric field.

Eq.~(\ref{125a}) can be written as a biquadratic algebraic equation of $r$. Solving this equation for $r=r(F^2)$ and substituting the expression into Eq.~(\ref{125b}), we find $K$ can be be explicitly expressed as the function of $F^2$.

\section{Discussion and conclusions}\label{sec:8}
By using the non-linear electromagnetic field theories, Ayon-Beato and Garcia proposed a powerful method to generate
black hole solutions \cite{bea:1998}. The method was successfully applied to build up the modified Reissner-Nordstr$\ddot{\textrm{o}}$m solution and the Bardeen solution. It was later generalized by Dymnikova \cite{dym:2015} for spherical and static solutions. More researches can be
found in \cite{noj:2017,eli:2002,dym:2004,ans:2007,ans:2008,joh:2013,dym:2015,cul:2017,ma:2015,kun:2015,pra:2015,rod:2016,fan:2016,chi:2017,fer:2017,pon:2017,cai:2004}.

In this paper, we generalize the black hole and cosmos spacetime to the scenarios of multiple horizons and multiple singularities. They are solutions of general relativity with the non-linear electromagnetic field. The solutions have regular event
horizons, approaches asymptotically flat or non-flat, and all its singularities
are concealed by the horizons. Since the spacetimes are all static and spherically symmetric, we could
calculate the surface gravity (see \cite{mae:2010} for details) and such that the temperatures.
For example, we find the 4-horizon black hole temperature on the
outermost horizon by the surface gravity as
\begin{eqnarray}
T=\frac{\kappa}{2\pi}=\frac{1}{4\pi}\frac{\left(a_4-a_1\right)\left(a_4-a_2\right)\left(a_4-a_3\right)}{a_4^4}\;.
\end{eqnarray}
We have assumed $a_1<a_2<a_3<a_4$. But when $a_3=a_4$, we have $T=0$ which is for the extreme black hole case. It is therefore interesting to discuss the thermodynamical properties since we can define the entropy and temperature
in the spacetime. As an example, Ref.~\cite{hendi:2015} investigates the thermodynamic instability problem of higher dimensional topological black holes in the presence of nonlinear electrodynamics. The extreme 4-horizon black holes have the vanishing surface gravity which means their gravitational attraction is exactly balanced by the Coulomb repulsion. So, motivated by the Kastor-Traschen solution, can we construct multiple 4-horizon black holes in the de Sitter universe? This is an open question. On the other hand, the orbits of test particles in the background of these spacetimes may be interesting since the causal structure of these spacetimes is rich.
Continuum spectrum from black hole accretion disc holds enormous information regarding the black hole physics \cite{ban:2017}. The signatures sculptured on these black hole might be distinct by the observations of continuum spectrum. Finally, by calculating the perturbations to the metric or the quasinormal modes of test fields, we hope the deviation of these black hole solutions may be potentially tested by future measurements of gravitational waves.

\section*{Acknowledgments}
We are grateful to the referees for the expert suggestions which have improved the paper significantly.
This work is partially supported by China Program of International ST Cooperation 2016YFE0100300
, the Strategic Priority Research Program ``Multi-wavelength Gravitational Wave Universe'' of the
CAS, Grant No. XDB23040100, the Joint Research Fund in Astronomy (U1631118), and the NSFC
under grants 11473044, 11633004, 11773031 and the Project of CAS, QYZDJ-SSW-SLH017.

\begin{table*}[h]
\begin{center}
\begin{tabular}[b]{cccc}
 \hline \hline
 \;\;\;\; $a_4$ \;\;\;\; & \;\;\;\; $\omega\ \ \ (l=1)$\;\;\;\;  & \;\;\;\;  $\omega \ \ \ (l=2)$\;\;\;\;
 & \;\;\;\; $\omega \ \ \ (l=3)$ \;\;\;\; \\ \hline
\\
4.0& \;\;\;\;\;0.085240-0.018464i\;\;\;\;\;  & \;\;\;\;
0.141767-0.017561i\;\;\;\;\;
 & \;\;\;\;\;0.198349-0.017305i\;\;\;\;\;
 \\
 3.9&0.086245-0.018599i&0.143464-0.017689i&0.200732-0.017429i
 \\
  3.8&0.087263-0.018736i&0.145186-0.017821i&0.203150-0.017559i
 \\
  3.7&0.088304-0.018911i&0.146933-0.017957i&0.205604-0.017693i
 \\
  3.6&0.089358-0.019066i&0.148707-0.018104i&0.208093-0.017832i
 \\
  3.5&0.090424-0.019219i&0.150504-0.018254i&0.210616-0.017980i
 \\
  3.4&0.091508-0.019378i&0.152325-0.018405i&0.213173-0.018130i
 \\
 3.3&0.092608-0.019555i&0.154172-0.018571i&0.215763-0.018290i
 \\
 3.2&0.093721-0.019715i&0.156042-0.018740i&0.218385-0.018458i
  \\
3.1&0.094851-0.019920i&0.157935-0.018919i&0.221037-0.018630i
\\
3.0&0.095997-0.020132i&0.159849-0.019106i&0.223721-0.018819i
\\
\hline \hline
\end{tabular}
\end{center}
\caption{The fundamental ($n=0$) quasinormal frequencies of scalar
field for the 4-horizon black hole for
$l=1$, $2$, $3$.}
\end{table*}

\newcommand\ARNPS[3]{~Ann. Rev. Nucl. Part. Sci.{\bf ~#1}, #2~ (#3)}
\newcommand\AL[3]{~Astron. Lett.{\bf ~#1}, #2~ (#3)}
\newcommand\AP[3]{~Astropart. Phys.{\bf ~#1}, #2~ (#3)}
\newcommand\AJ[3]{~Astron. J.{\bf ~#1}, #2~(#3)}
\newcommand\APJ[3]{~Astrophys. J.{\bf ~#1}, #2~ (#3)}
\newcommand\APJL[3]{~Astrophys. J. Lett. {\bf ~#1}, L#2~(#3)}
\newcommand\APJS[3]{~Astrophys. J. Suppl. Ser.{\bf ~#1}, #2~(#3)}
\newcommand\JHEP[3]{~JHEP.{\bf ~#1}, #2~(#3)}
\newcommand\JMP[3]{~J. Math. Phys. {\bf ~#1}, #2~(#3)}
\newcommand\JCAP[3]{~JCAP {\bf ~#1}, #2~ (#3)}
\newcommand\LRR[3]{~Living Rev. Relativity. {\bf ~#1}, #2~ (#3)}
\newcommand\MNRAS[3]{~Mon. Not. R. Astron. Soc.{\bf ~#1}, #2~(#3)}
\newcommand\MNRASL[3]{~Mon. Not. R. Astron. Soc.{\bf ~#1}, L#2~(#3)}
\newcommand\NPB[3]{~Nucl. Phys. B{\bf ~#1}, #2~(#3)}
\newcommand\CMP[3]{~Comm. Math. Phys.{\bf ~#1}, #2~(#3)}
\newcommand\CQG[3]{~Class. Quantum Grav.{\bf ~#1}, #2~(#3)}
\newcommand\PLB[3]{~Phys. Lett. B{\bf ~#1}, #2~(#3)}
\newcommand\PRL[3]{~Phys. Rev. Lett.{\bf ~#1}, #2~(#3)}
\newcommand\PR[3]{~Phys. Rep.{\bf ~#1}, #2~(#3)}
\newcommand\PRd[3]{~Phys. Rev.{\bf ~#1}, #2~(#3)}
\newcommand\PRD[3]{~Phys. Rev. D{\bf ~#1}, #2~(#3)}
\newcommand\RMP[3]{~Rev. Mod. Phys.{\bf ~#1}, #2~(#3)}
\newcommand\SJNP[3]{~Sov. J. Nucl. Phys.{\bf ~#1}, #2~(#3)}
\newcommand\ZPC[3]{~Z. Phys. C{\bf ~#1}, #2~(#3)}
\newcommand\IJGMP[3]{~Int. J. Geom. Meth. Mod. Phys.{\bf ~#1}, #2~(#3)}
\newcommand\IJMPD[3]{~Int. J. Mod. Phys. D{\bf ~#1}, #2~(#3)}
\newcommand\GRG[3]{~Gen. Rel. Grav.{\bf ~#1}, #2~(#3)}
\newcommand\EPJC[3]{~Eur. Phys. J. C{\bf ~#1}, #2~(#3)}
\newcommand\PRSLA[3]{~Proc. Roy. Soc. Lond. A {\bf ~#1}, #2~(#3)}
\newcommand\AHEP[3]{~Adv. High Energy Phys.{\bf ~#1}, #2~(#3)}
\newcommand\Pramana[3]{~Pramana.{\bf ~#1}, #2~(#3)}
\newcommand\PTP[3]{~Prog. Theor. Phys{\bf ~#1}, #2~(#3)}
\newcommand\APPS[3]{~Acta Phys. Polon. Supp.{\bf ~#1}, #2~(#3)}
\newcommand\ANP[3]{~Annals Phys.{\bf ~#1}, #2~(#3)}

\end{document}